# The adaptive acquisition of single DNA segments drives metabolic evolution across *E. coli* lineages

Tin Yau Pang & Martin J. Lercher

**Abstract**

Even closely related prokaryotes show an astounding diversity in their ability to grow in different nutritional environments[1,2]. Mechanistically, this diversity arises predominantly through horizontal gene transfer, the exchange of DNA between individuals from different strains[3]. It has been hypothesized that complex metabolic adaptations – those requiring the acquisition of multiple distinct DNA segments – can evolve via selectively neutral intermediate steps[4,5]; an alternative explanation rests on the existence of intermediate environments that make each individual DNA acquisition adaptive[6]. However, it is unclear how important changing environments[6] are compared to neutral explorations of phenotype space[5]; more fundamentally, it is unknown what fraction of metabolic adaptations are indeed complex. Here, we use metabolic network simulations to show that all 3,363 phenotypic innovations observed in the evolutionary history of 53 *E. coli* strains arose through the acquisition of a single DNA segment; while we found no evidence for the contribution of selectively neutral processes, 10.6% of adaptations to previously unviable environments relied on the support of DNA acquisitions on earlier phylogenetic branches. 97.0% of all metabolic phenotypes accessible for the *E. coli* pan-genome could be bestowed on any ancestral strain by transferring a single DNA segment from one of the extant strains. These results demonstrate an amazing ability of the *E. coli* lineage to quickly adapt to previously inaccessible environments through a single DNA acquisition, an ability likely to be mirrored in other clades of generalist bacteria.

**Introduction**

With the advent of high-throughput DNA sequencing, comparative genomics replaced the comparison of phenotypes as the basis for understanding evolution and natural selection. However, it is the phenotype that natural selection acts upon; to fully appreciate the patterns and driving forces of adaptation, we need to link genotypes to phenotypes both on the genomic scale and on an evolutionary timescale. Bacterial metabolism is arguably the most promising model system for such an endeavour: the ability to metabolize different nutrient sources always remains an essential determinant of bacterial fitness[2], and flux balance analysis (FBA) has been established as a robust and reliable modeling framework for the prediction of this ability[7].

Experimental growth data indicates that genetic distance, based on multilocus sequence typing data, is a weak indicator of how similar two *E. coli* strains are in terms of the carbon sources they can metabolize[1]. Similarly, a computational analysis of automatically generated metabolic models for hundreds of species suggested that within-species phenotypic divergence is almost instantaneous, while divergence between genera is gradual or "clock-like"[2]. How can within-species divergence be so much faster than between-species divergence? The answer likely lies in frequent recombination between bacterial strains belonging to the same species, a process that facilitates the transfer of genes

(horizontal gene transfer, HGT)[8]. This process effectively decouples phenotypic divergence, dominated by HGT of metabolic pathways[2], from genome sequence divergence, dominated by mutations accumulating along lines of vertical inheritance. Thus, the *E. coli* pan-genome acts like a repository from which individual strains can add metabolic tools to their existing toolboxes[9,10], a process independent from vertical inheritance.

Horizontally transferred genes that do not provide fitness benefits are likely to be quickly lost, not least because of a mutational bias towards deletions in bacterial genomes[11]. This suggests that successful HGT events, *i.e.*, those that their traces in extant genomes, were individually adaptive. This would impose a strong barrier for the emergence of complex phenotypes that require the acquisition of multiple genes located on distinct genomic regions, as DNA segments successfully transferred between *E. coli* strains practically never exceed 30kb in length[12].

Did ancient strains of the *E. coli* lineage find a way to circumvent this barrier? It has been proposed that complex metabolic adaptations may evolve via a neutral exploration of phenotype space[4,5], hypothesizing that "many additions of individual reactions to a metabolic network will not change a metabolic phenotype until a second added reaction connects the first reaction to an already existing metabolic pathway"[5]. However, no empirical data from bacterial metabolism supports this scenario; bacterial genomes appear compact and almost devoid of non-functional DNA sequences[11].

An alternative explanation for the emergence of complex adaptations was put forward by Szappanos *et al.*[6], suggesting that metabolic complexity may arise through successive non-complex adaptations to changing environments[6]. However, the relative roles of simple adaptations (requiring only a single DNA acquisition) and complex adaptations in bacterial evolution are currently unknown. What proportion of metabolic innovations in the *E. coli* clade were complex, *i.e.*, required multiple independent HGT events? Did such multiple DNA acquisitions occur simultaneously, or were they spread over long evolutionary time spans, suggesting a role for successive adaptations to "stepping stone" environments? And, more fundamentally: how adaptable are *E. coli* strains, *i.e.*, how many independent DNA acquisitions are typically required to enable growth in a random, previously unviable environment?

## Results

Here, we examine these questions by simulating the metabolism of extant and ancestral *E. coli* strains, enabling us to link genomic and phenotypic evolution in this clade. We analyzed 53 *E. coli* and *Shigella* strains (Supplementary Figure S1, Supplementary Table S1), encompassing commensal as well as intestinal and extraintestinal pathogenic strains[13]. Because the *Shigella* strains are nested within the *E. coli* phylogeny[14], we subsume them in the following under the term *E. coli*. For theses 53 strains, previous publications provided genome-scale metabolic models[13] as well as a reliable phylogeny of vertical inheritance together with the likely presence and absence of genes in the ancestral strains[12]. We reconstructed the metabolic networks of the 52 ancestral strains (Materials and Methods) and performed flux balance analysis (FBA) on the ancestral and extant networks, testing their ability to grow in 200,000 randomly generated nutritional environments as well as in 2,418 environments used in previous simulations of *E. coli* metabolic adaptation[6] (Supplementary Table S5). 30 extant and 46 ancestral networks were each able to grow in more than 40,000 environments (Table S1); the remaining networks produced biomass in only a small minority of the tested environments and were excluded from further analyses.

As expected, strains that diverged very recently (amino acid sequence divergence <0.01%) tend to grow in the same environments. Beyond those nearly identical strains, however, we find that phenotypic similarity is independent of the amino acid sequence divergence of two strains (Figure 4A; Spearman's $\rho$=0.0052, $P$=0.88), an observation that agrees with earlier observations[1,2].

To study the potential spread of phenotypic innovations through HGT within the *E. coli* pan-genome, we developed a model of functional HGT based on the extant and ancestral metabolic networks. We merged the metabolic network of the *E. coli* strains into a super-network, performing FBA to identify all possible metabolic phenotypes across the *E. coli* pan-genome. The pan-genome-scale super-network showed new phenotypes in comparison to each of the 30 extant genomes and was able to produce biomass in many environments inaccessible to any of the extant strains. For each extant genome, we identified the minimal number of 30kb segments from other extant genomes that would have to be jointly transferred to bestow a new phenotype (the ability to grow in a previously inaccessible environment) or a strongly enhanced phenotype (at least a doubling of biomass yield) found for the super-network; 30kb represent an upper size limit for DNA segments successfully acquired via HGT by *E. coli* strains[12]. Based on these simulations, we expect that only 2.4% of new phenotypes and 7.4% of strongly enhanced phenotypes require the acquisition of multiple DNA segments, *i.e.*, should be classified as complex adaptations (Figure 2).

If a strain can grow in a certain environment, how likely is it to be able to confer this ability to another strain through the transfer of a single DNA segment? To test this, we analyzed all potential donor-recipient pairs of extant *E. coli* strains, examining all environments in which the potential donor but not the recipient can grow. For most donor-recipient pairs, all such differential phenotypes can be bestowed on the recipient through the transfer of a single DNA segment of the donor; on average, 98.8% of growth abilities can be transferred in this way (Figure 4B). While there is a weak negative correlation between phenotype transferability and amino acid divergence (Spearman's $\rho$=-0.24, $P$<10$^{-12}$), the majority of phenotypes can be transferred with a single DNA segment even for the most divergent donor-recipient pairs (Figure 4B).

The probability that a DNA segment that had conferred a given phenotype to one strain is able to confer the same phenotype to another strain decreases slightly with genomic sequence divergence (Spearman's $\rho$=-0.20, $P$<10$^{-8}$), but remains above 90% regardless of sequence divergence. The potential utility of HGT, however, does not depend on sequence divergence: the probability that the transfer of a random DNA segment not currently present in the recipient leads to *any* phenotype innovation remains at around 23.8% averaged over all possible DNA segments exchanged between two strains (Figure 4C; Spearman's $\rho$=-0.0064, $P$=0.85).

How does this picture of potential phenotypic innovation through HGT compare to the phenotypic changes actually observed throughout the *E. coli* evolutionary history? From FBA simulations on the extant and ancestral metabolic networks, we identified phenotype innovations in the derived node of each branch of the rooted *E. coli* phylogeny (relative to its direct ancestor). Here, we consider phenotype innovations as (i) the ability to grow in additional metabolic environments (new metabolic phenotypes) and (ii) the ability to at least double the biomass yield in one or more metabolic environments (strongly improved phenotypes). We performed parsimonious FBA (pFBA)[15] to identify the metabolic reactions contributing to the phenotype innovations. Reactions present at the derived but not the ancestral node of a branch must have been acquired horizontally, and we assigned the genes encoding the corresponding enzymes to a minimal set of transferred DNA segments at most 30kb long[12].

We were able to identify at least one new phenotype for 55.8% and one strongly enhanced phenotype for 58.9% of the DNA segment acquisitions observed along the *E. coli* phylogeny (Figure 1). All observed metabolic innovations were achieved through the horizontal acquisition of a single DNA segment (Figure 2), a result that is in stark contrast to speculations about a neutral exploration of phenotypic space in metabolic evolution[5].

Complex innovations may be expected to be rare due to a high similarity between the metabolic systems of different *E. coli strains*, combined with the organisation of functionally related genes into operons – we have already seen that almost all potential phenotypic innovations from the *E. coli* pan-genome can be acquired via HGT of a single DNA segment (Figure 2). But complex innovations are still possible, both as alternatives to single-segment adaptations and in the rare cases where multiple DNA segments are indeed necessary to evolve a new phenotype. That we did not find a single complex adaptation among the 3,323 observed *E. coli* adaptations demonstrates that metabolic evolution is unlikely to be a predominantly neutral process ($P<2.2\times10^{-16}$, 2-sided binomial test). We thus conclude that only DNA segments that are individually adaptive have spread through *E. coli* populations, and that complex phenotypes rarely or never evolve through neutral evolution within the *E. coli* pan-genome.

It has been proposed that exaptation plays an important role in metabolic evolution, i.e., adaptive metabolic innovations accessible through a single HGT event may serve as stepping stones towards the later establishment of more complex metabolic features in another environment[6]. Over the time scale of individual branches of the *E. coli* phylogeny, such exaptations appear to have played at most a minor role in *E. coli* metabolic evolution (Figure 2). What about larger evolutionary time frames? For each phenotypic innovation that evolved on the branch leading to an extant or ancestral strain, we identified the number of DNA segments required for this phenotype that were acquired since the last common ancestor of all analyzes *E. coli* strains. As shown in Figure 3, 10.6% of new phenotypes and 19.0% of improved phenotypes indeed relied on combining multiple DNA segment transfers since the origin of the *E. coli* clade, using adaptations on earlier branches as evolutionary stepping stones. As an example, Supplementary Figure S3 shows the pathway to metabolize D-Allose and D Glycerate 2 phosphate of fructoselysine metabolism in an O157 strains, which contains reactions acquired through HGT on two different branches of the phylogeny.

**Discussion**

We observed that there is no molecular clock for phenotypic divergence within species (as observed on larger evolutionary scales[2]. Thus, *E. coli* strains have a well-filled metabolic "toolbox"[9,10]: the vast majority of phenotypic innovations that could potentially be acquired from the *E. coli* pan-genome are accessible through a single HGT event.

Our results provide a comprehensive picture of metabolic evolution across the *E. coli* pan-genome. Within the *E. coli* species, where recombination and, consequently, HGT are frequent[16], sequence similarity does not influence phenotypic similarity. This contrasts with observations of long-term metabolic evolution, where phenotypic divergence appears to increase with evolutionary distance[2]. The lack of phenotypic divergence within *E. coli* may be linked to our observation that even for the most diverged genomes, there is still a 99% chance that a given phenotype of one strain can be transferred to the other with a single DNA segment.

While sequence divergence thus does not impose a barrier to phenotype transfer within the *E. coli* pan-genome, the sequence divergence between two potential recipients is predictive of whether they will acquire the same new phenotypes. This emphasizes that while all *E. coli* strains possess a well-filled toolbox[9,10], these contain different tools (*i.e.*, enzymes and transporters): depending on which tools are already present, the same set of additional tools may facilitate different phenotypes. The well-filled toolboxes of *E. coli* strains also partly explain why all possible new phenotypes can be acquired with a single HGT event. Thus, contrary to earlier suggestions[17], neutral evolution is neither necessary nor is it observable across *E. coli* strains.

The prevalence of ready-to-transfer phenotypes is important for our understanding of bacterial pan-genomes. Our dataset contains strains from five major *E. coli* sub-groups (A, B1, B2, D, E), yet the chance of successful phenotype transfer remained at almost 99% even across subclades. Thus, the adaptive utility of DNA segments is not restricted to HGT within *E. coli* subclades. We conclude that metabolic adaptation within the wider *E. coli* pan-genome is an extremely efficient process, where the HGT of a single DNA segment is sufficient to facilitate the jump of any non-auxotrophic strain from its home environment to virtually any other reachable environment.

It has been suggested that metabolic evolution may often proceed through adaptations to "intermediate" environments that act as evolutionary stepping stones, providing reactions that can later be exapted for additional adaptations[6]. That all possible phenotypic innovations can in principle be acquired by any *E. coli* strain through acquisition of a single DNA segment already suggests that exaptation may play only a limited role in *E. coli* evolution. We do, however, still observe the contribution of exaptations to 10.6% of new phenotypes and 19.0% of strongly enhanced phenotypes, emphasizing that such stepwise metabolic niche expansion did indeed make a significant contribution even to the evolution of the generalist species *E. coli*.

**Materials and Methods**

*Universal GPR rules*

We start from the previously published metabolic network reconstructions of 53 extant *E. coli* strains[13]; we excluded the two strains SE11 and 55989, as their position on the phylogeny describing vertical descent among *E. coli* strains was uncertain[12]. Note that we use the term *E. coli* as including all Shigella strains, as these are nested within the *E. coli* clade[12].

Some of the gene-protein-reaction (GPR) rules provided in the xml files of the metabolic networks reconstructed by *Monk et al.*[13] are inconsistent between *E. coli* strains. Thus, we first defined a universal set of GPR rules. For this, we merged all alternative GPRs from different strains by a logical OR; *e.g.*, if a reaction is catalyzed by an enzyme complex encoded by a set of orthologous gene families in one strain but by a gene product from a single gene family in another strain, we assumed that the reaction can proceed as long as either the enzyme complex members or the single enzyme are encoded in the genome. We allowed a reaction to proceed without genomic evidence (an empty GPR) only if the GPR was empty in all 53 metabolic models. We applied the universal GPR rules to the 53 extant strains, replacing the existing GPR rules with the universal GPR rules. If one of the alternative enzymes or enzyme complexes linked by logical ORs is not fully supported by a strain's genome, this part of the GPR is removed; if none of the enzymes or enzyme complexes linked by OR is supported by the genome, the reaction is removed from the metabolic model.

### *Reconstruction of ancestral metabolic networks from ancestral genomes*

We applied these universal GPR rules to the gene content of all ancestral strains, obtained from *Pang et al.*[12]. For each ancestral strain, we formed a metabolic network reconstruction out of all reactions supported by its genome based on the universal GPR rules; we always include all reactions with an empty GPR rule. Reaction reversibilities and biomass reactions are universal across the metabolic reconstructions published in *Monk et al.*[13], and were copied to the ancestral metabolic reconstructions. We used the standard biomass reaction (Ec_biomass_iJO1366_core_53p95M) in all simulations. All models also contained a non-growth related maintenance energy consumption term (a flux of at least 3.15 through the ATPM reaction, which converts ATP to ADP). Supplementary Table S1 shows the number of genes, metabolic genes, and metabolic reactions in each of the extant and ancestral strains. Supplementary Table S3 shows the compiled universal GPR rules.

### *Reconstruction of genes transferred horizontally on DNA segments*

Based on the ancestral genomes provided by Pang *et al.*[12], we reconstructed the gene acquisitions via horizontal gene transfer (HGT) for each branch of the phylogeny provided in the same publication. We then identified sets of genes that were likely co-transferred on a single DNA segment using a greedy algorithm; this inference was based on the observation that DNA segments acquired via HGT by *E. coli* strains are up to 30kb in length[12].

Let **S** be the set of all orthologous gene families ("genes") acquired along a given branch. Let **E** be the set of of extant descendants of this branch. We then carried out the following iterative algorithm to group the horizontally transferred genes into subsets of genes that were co-transferred on a single DNA segment:
1. identify the start positions of every orthologous gene family in **S** in the genomes in **E**;
2. pick a random orthologous gene family $g_A$ in **S** and put it into a new set **P**;
3. for each gene $g$ in **S** not included in **P** (□$g \in$ **S\P**), count the number of extant genomes in **E** where $g$ together with all genes in **P** is contained within a 30kb segment of the genome;
4. pick the one gene $g$ in **S** not included in **P** ($g \in$ **S\P)** supported by the highest segment count, and add it to **P**; if multiple genes are supported equally, pick one randomly;
5. repeat steps 3-4 until no more genes can be added to **P;**
6. the genes in **P** are then grouped into a single HGT event; these genes are removed from **S**;
7. Repeat steps 2-6 to reconstruct another HGT event until **S** is empty (*i.e.*, all horizontally acquired genes have been assigned to a DNA segment).

Supplementary Table S2 shows our reconstruction of HGT segments and their associated genes. To test the reliability of our reconstruction, we repeated our algorithm a thousand times. We observed that a set of genes grouped into a DNA segment in one realization has a 99.3% chance to also occur as a co-transferred set in another realization.

### *Quantifying the metabolic phenotypes of extant and ancestral E. coli strains in different nutritional environments*

We applied flux balance analysis (FBA) to calculate the biomass production rate of each extant and ancestral metabolic model across 202,418 different nutritional environments. 2,418 of these nutritional environments are based on *Szappanos et al.*[6]; *Szappanos et al.* tested 1,776 nutritional environments, of which 1,209 nutritional environments remained after removing those environments that contained exchange reactions not found in the metabolic models of *Monk et al.*[13]. We doubled this set of 1,209 aerobic nutritional environments by adding corresponding anaerobic environments (removing oxygen uptake).

We further generated 100,000 different random nutritional environments, constructed by choosing a random source each of carbon, nitrogen, phosphorus, and sulphur (see Supplementary Table S4 for the lists of compounds as possible carbon, nitrogen, phosphorus, and sulphur sources); to each environment, we added 12 other elements or compounds: calcium (ca2), chloride (cl), copper (cu2), cobalt (cobalt2), iron (fe3), water (h2o), potassium (k), magnesium (mg2), manganese (mn2), molybdate (mobd), zinc (zn2) and nickel (ni2). These 100,000 nutritional environments are anaerobic. We added the corresponding aerobic environments, resulting in a total of 200,000 random environments. Supplementary Table S5 lists the nutrients in each of the 202,418 environments tested. When performing FBA simulations, we set the lower bound of all exchange reactions for nutrients present in the tested environment to -10; the lower bound of all other exchange reactions is set to 0.

We used the sybil package in R[18] to perform FBA simulations on the 53 extant and 52 ancestral strains across the 202,418 nutritional environments. Maximal biomass reaction rates $\leq 10^{-6}$ were interpreted as an inability of the strain to grow in the tested environment. Supplementary Figure S4 shows the distribution of the number of viable environments per strain. 76 of the metabolic models (46 ancestral and 30 extant strains) can grow in at least 5,000 of the tested environments, while the remaining 29 models can produce biomass in at most 0.5% of the tested environments; 26 models actually produce biomass in none of the tested environments. These 29 strains were either highly metabolically specialized or had incomplete metabolic reactions (*e.g.*, because reactions added by Monk *et al.*[13] without genomic evidence were removed in our application of the universal GPR rules); thus, we excluded these strains from further analysis, presenting results only for the remaining 76 strains (30 extant and 46 ancestral).

### *Phenotypic and genomic distances between strains*

We defined the phenotypic distance of a pair of strains as 1-*J*, where *J* is the Jaccard index of the subsets of nutritional environments (out of the 202,418 tested environments) in which each of the two strains can grow according to the FBA simulations. We defined the genomic distance between two strains as the amino acid sequence distance calculated per amino acid site from the concatenated global alignment of 1,334 core genes[12].

*Assessing phenotype innovations and corresponding DNA segment transfers in the real data and in a horizontal gene transfer (HGT) model*

To evaluate the phenotype innovations conferred by DNA segment transfer events in the real data, we applied FBA to the metabolic networks at the beginning and end of the phylogenetic branch on which one or more transfers occurred. To guard against numerical imprecisions, we used a threshold of $10^{-6}$ to distinguish "non-zero biomass production" (growth) from "zero biomass production" (non-growth). We compared the biomass production of the ancestral and derived strains in each of the 202,418 environments, and selected those environments that have phenotype innovations that fall into one of two phenotype innovation types:

   I. New phenotypes, where the derived strain can grow in the tested environment while the ancestral strain cannot .
   II. Strongly enhanced phenotypes, where both ancestral and derived strains can grow in the tested nutritional environment, and the derived strain has a biomass flux at least twice as as that of the  ancestral strain.

To study more generally how HGT can lead to phenotype innovations, we developed a model based on the 30 non-auxotrophic extant strains. In each strain, we identified all non-universal metabolic genes, with their genomic position represented by the transcription start sites. We then searched for all possible gene combinations that can be transferred within a 30kb DNA segment and that are not a subset of another set (30kb is the upper limit of the distance distribution of DNA segments acquired in a single HGT event by *E. coli* strains[12]). The sets of genes defined in this way thus correspond to all metabolically relevant potential horizontal gene transfers of DNA segments (see Supplementary Table S6 for these segments). For each segment, we modeled its transfer to each of the other 29 non-auxotrophic extant strains that misses at least some of its genes, performed FBA to evaluate the biomass flux in each of the 202,418 nutritional environments, and enumerated the two types of segment transfer effects.

*Calculating the phenotype-transferability between a pair of donor and recipient*

To test the extent to which pairs of strains can benefit from HGTs, we examined every pair of extant non-auxotrophic strains. Given a donor and a recipient, we considered the phenotypes present in the donor but not in the recipient ("unique phenotypes"). A phenotype is single-segment-transferrable if there exists a DNA segment in the donor such that the transfer of the segment to the recipient leads to the transfer of the phenotype. The phenotype transferability of a donor-recipient pair is defined as the proportion of transferrable phenotypes out of all unique phenotypes of the pair.

*Calculating the success rate for segment-transfer to result in phenotype-transfer*

We also calculated the probability that a segment bestows any new phenotype on its recipient, regardless of whether that phenotype was present in the donor. For a pair of extant non-auxotrophic donor and recipient, we calculated the fraction of donor DNA segments containing metabolic genes not present in the recipient that give rise to at least one new phenotype in the other strain.
    We also tested if the transfer of the same DNA segment into two recipients leads to the same phenotype innovation. For a pair of extant non-auxotrophic strains A and B, we

considered all environments in which both A and B cannot grow. We then examined all DNA segments of other strains that can bring at least one new metabolic gene to each of their genomes and simulated the consequences of their acquisition by both A and B using FBA. Based on these results, we calculated the probability for a segment to give rise to the same phenotype in both A and B.

***Estimating the minimal number of DNA segments required for a phenotype innovation***

To estimate the number of DNA segments required for a phenotype innovation observed in the real data, we first performed parsimonious FBA (pFBA)[15] on the metabolic network of the derived strain of the branch in the environment for which a phenotype innovation occurred. pFBA reports the reactions most likely involved in the nutritional phenotype; we identified the set of active reactions that have been transferred into the branch. Based on the previously identified transferred DNA segments (Supplementary Table S2), we then identified the number of segments involved in the phenotype innovation. As pFBA only approximately describes the physiology of biological systems[15], this procedure might inflate the number of segments contributing to the phenotype innovation. Hence, when pFBA suggested that there are more than 1 segments contributing to a phenotype innovation, we simulated transferring all possible combinations of those predicted segments to the ancestral network to determine the minimal number of segments able to bestow the observed phenotype innovation.

As a reference, we also investigated the number of segments required for any phenotype innovation in the HGT model. We first constructed the metabolic supermodel that contains all reactions from the 30 extant non-auxotrophic strains, and performed pFBA on the 202,418 nutritional environments to obtain the biomass flux and reaction pathways in the supermodel. We compared the biomass flux of different environments of the supermodel with that of each non-auxotrophic extant strain, and identified the potential phenotype innovations. pFBA performed on the supermodel predicts the reactions required by a nutritional phenotype, and those reactions that are not present in the considered extant strain must be acquired through HGT. We searched for the minimum number of segments in the extant genomes that can provide those extra reactions (see **Supplementary Table S6** for potential segments). If the pFB-based method predicts multiple segments, we transferred each segment found on the extant strains (**Supplementary Table S6**) to the recipient strain to test if a single segment transfer can already result in the phenotype innovation.

In the analysis of the real data, we identified 1998 new phenotypes (type I) and 1325 strongly enhanced phenotypes (type II), and found that none of them required more than a single DNA segment transfer to arise. In the HGT model, the fraction of potential innovations that require multiple DNA segment transfers was 0.0239 for new phenotypes (type I) and 0.0739 for enhanced phenotypes (type II). We applied a single-sided binomial test to examine the hypothesis that the chance of multi-segment phenotype innovations in the real data is lower than predicted by the HGT model ($P<10^{-15}$ for both types).

***Identifying exaptations of earlier HGT events for later phenotype innovations***

In order to assess the role of exaptation in *E. coli* metabolic adaptations, we identified phenotype innovations that utilized enzymes or transporters transferred into the lineage on previous branches of the phylogeny. For each ancestral and extant metabolic network, we used the pFBA results calculated above to identify the active reactions involved in any newly evolved phenotype innovation (types I and II). Each of these reactions was either (i) inherited from the most recent common ancestor (MRCA) of all examined strains—the root node of the phylogeny—, (ii) acquired via HGT on the preceding phylogenetic branch—reactions directly responsible for the innovation—, or (iii) acquired via HGT on an earlier branch. The reactions in category (iii) were thus acquired much earlier and were exapted for the considered phenotype. We identified the corresponding minimal number of DNA segments gained via HGT on earlier branches (see Supplementary Table S2 for the segments transferred into different branches).

**Figures**

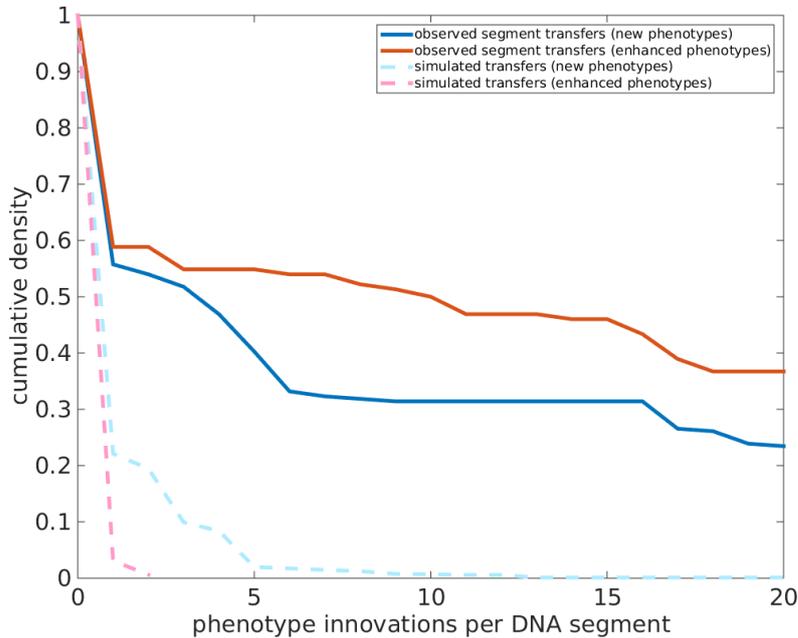

**Figure 1.** The successful horizontal transfer of DNA segments lead much more often to phenotypic innovations than expected for randomly chosen segments found in extant *E. coli* strains. Cumulative distribution of the number of phenotypic innovations per DNA segment.

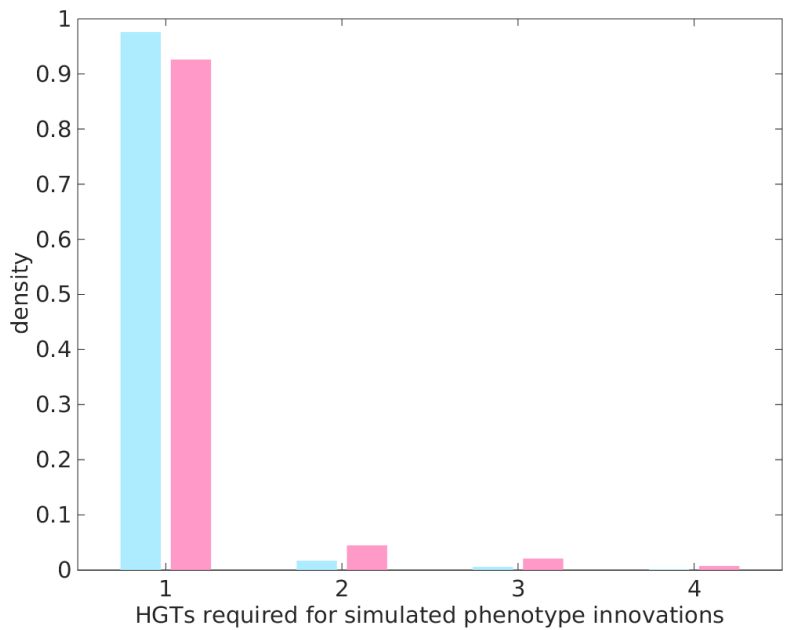

**Figure 2.** While every phenotype innovation observed in the *E. coli* lineage was conferred by the transfer of a single DNA segment, alternative potential phenotype innovations would sometimes have required the acquisition of several distinct segments.

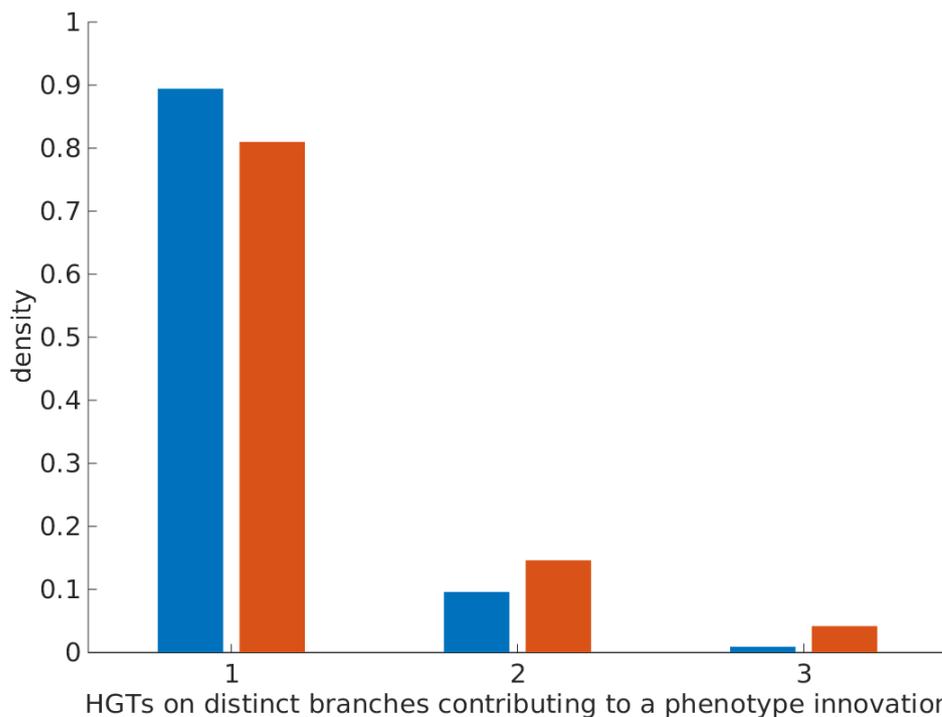

**Figure 3.** 10.6% of new and 19.0% of strongly enhanced phenotypes in the *E. coli* lineage evolved through 2-3 successive horizontal DNA acquisitions on distinct branches of the phylogeny. Note that every single one of these apparently "complex" phenotype innovations could instead have been bestowed through a single DNA segment presently found in one of the other extant *E. coli* strains.

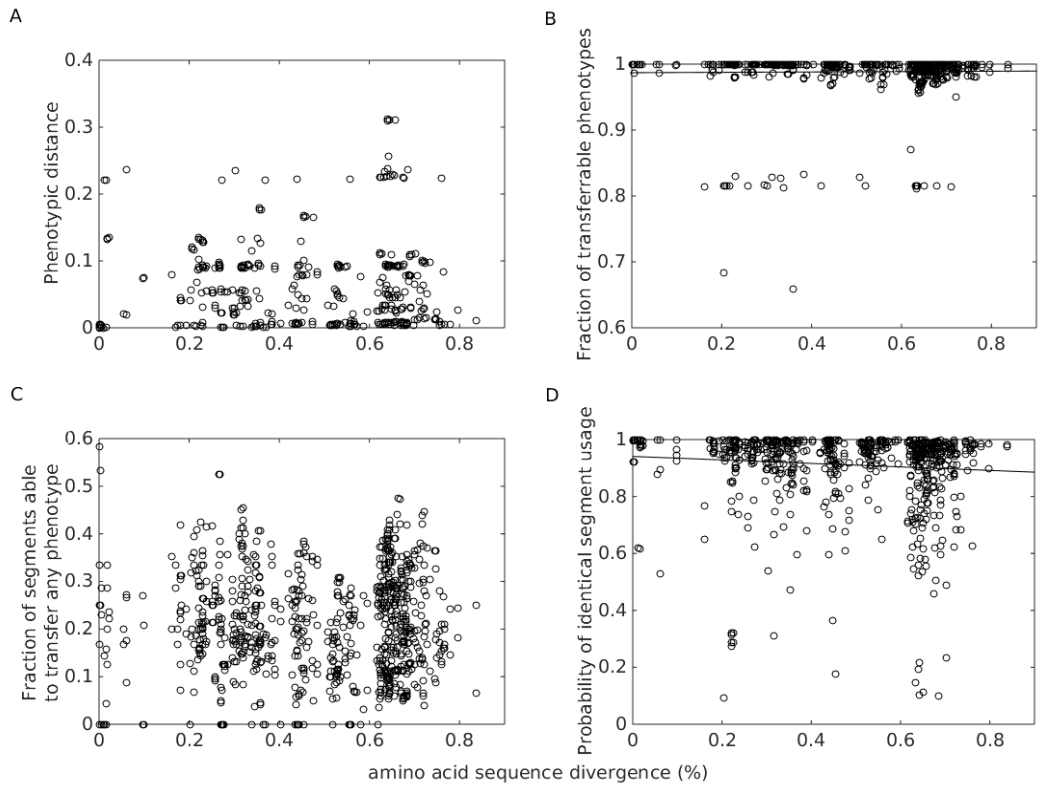

**Figure 4**. A: Phenotypic distance – defined as 1 minus the Jaccard index of the sets of environments in which two strains can grow – is independent of the genomic amino acid sequence divergence of the strains (Spearman's $\rho=0.0052$, $P=0.88$). B: Whether a phenotype can be transferred from one strain to another through a single DNA segment is independent of their amino acid divergence (Spearman's $\rho=-0.24$, $P<10^{-12}$). C: The probability for a segment to bestow any phenotype innovation on the recipient is independent of the donor-recipient divergence (Spearman's $\rho=-0.0064$, $P=0.85$). D: The probability that two distinct recipients acquire the same phenotype innovations from a given DNA segment depends on their amino acid sequence divergence (Spearman's $\rho=-0.20$, $P=10^{-9}$). Each circle in these figures represents the binning of 50 neighbouring data points.

## Supplementary Data

**Supplementary Table S1**. Details of the 53 strains, including their genbank ID, name, and numbers of genes, metabolic genes, metabolic reactions, and tested environments that support growth.

**Supplementary Table S2**. Reconstructed DNA segments containing enzymes and/or transporters that have been transferred into different branches of the *E. coli* phylogeny. Gene IDs in this table are mapped to locus tags in the genbank files through Supplementary Table S3 in *Pang et al.*[12].

**Supplementary Table S3**. Compiled universal gene-reaction (GPR) rules for inferring the metabolic network from the genomic presence and absence of orthologous gene families. Gene IDs in this table are mapped to locus tags in the genbank files through Supplementary Table S3 in *Pang et al.*[12]; reaction names follows that of the metabolic models in *Monk et al.* \cite{monk2013}.

**Supplementary Table S4**. Exchange reactions of the carbon, nitrogen, phosphorous and sulphur sources we used to generate the random nutritional environments.

**Supplementary Table S5**. Exchange reactions in each of the 202,418 nutritional environments tested in our study. The first 2,418 environments in the list come from *Szappanos et al.*[6], while the remaining 200,000 environments represent random combinations of carbon, nitrogen, phosphorous, and sulphur sources (see Supplementary Table S4).

**Supplementary Table S6**. Possible non-universal DNA segments observed in the extant genomes containing enzymes and/or transporters. Gene IDs in this table are mapped to locus tags in the genbank files through Supplementary Table S3 in *Pang et al.*[12].

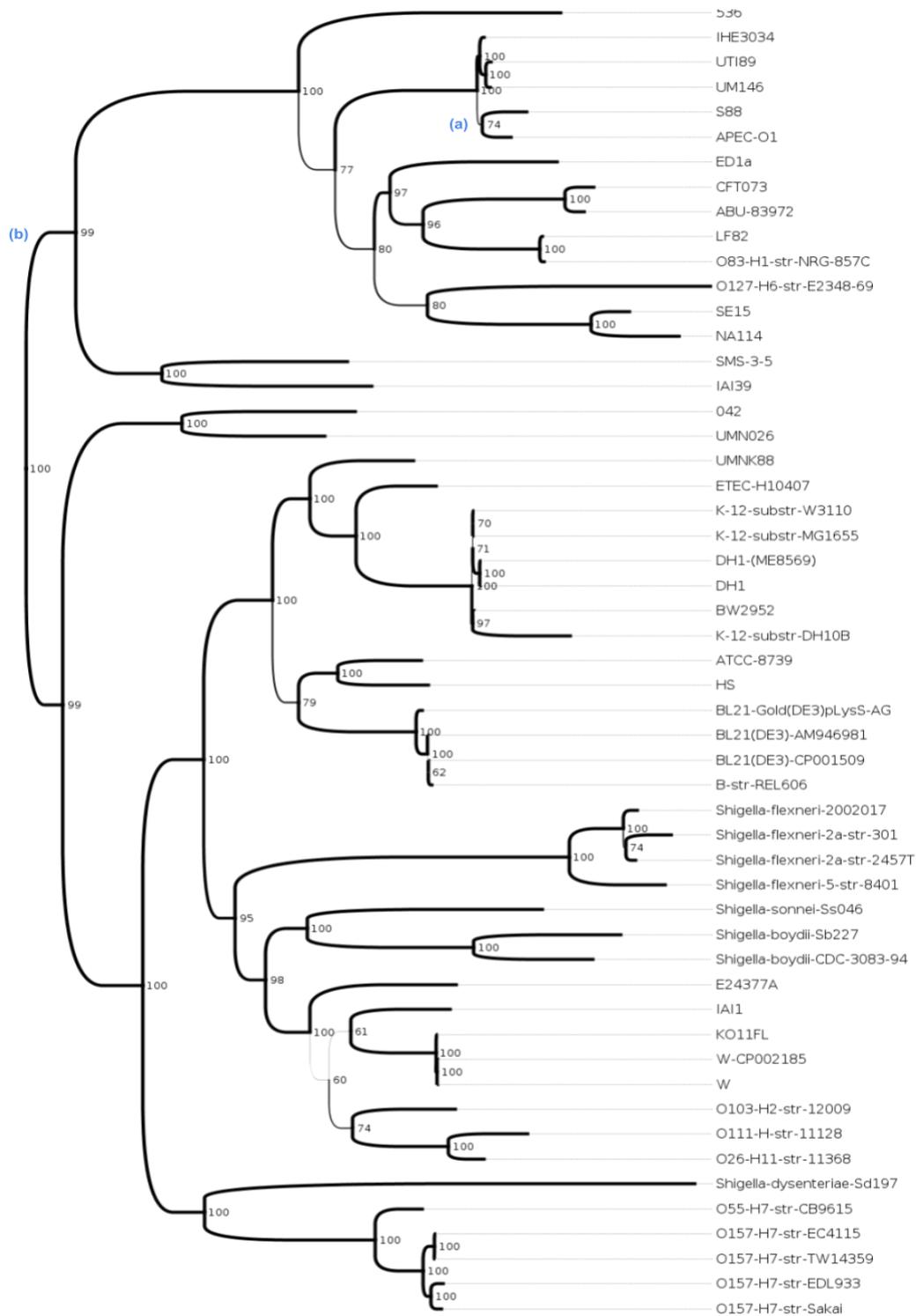

**Figure S1.** Phylogenetic tree describing vertical inheritance among the 53 *E. coli* strains (including *Shigella*) analysed in this work (redrawn from Figure S2 of *Pang et al.*[12]). Numbers to the right of branches (as well as branch line widths) reflect the maximum-likelihood bootstrap support values in percent. Reactions contributing to the complex phenotype innovation described in Supplementary Figure S3 were acquired via HGT on the branches marked with (a) and (b) to the left of the respective branches.

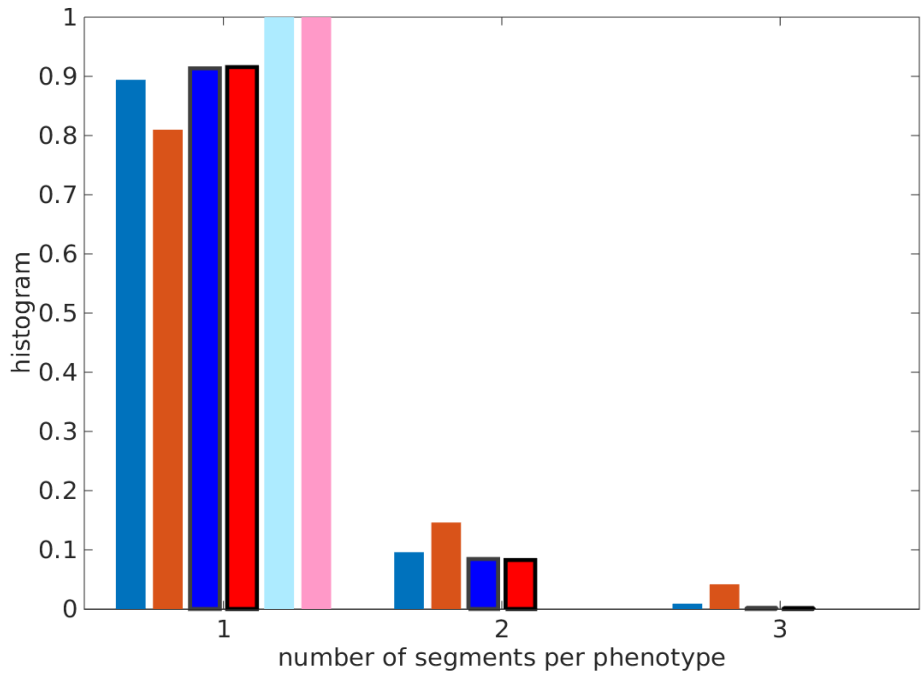

**Supplementary Figure S2**. 14% of phenotypic innovations evolved through 2-3 successive horizontal DNA acquisitions that occurred on distinct branches of the phylogeny. Histogram of the number of DNA segments required for new phenotypes. Blue: new phenotypes, red: strongly improved phenotypes; dark colors and no border: empirical data, dark colors and thick border: empirical data with transferred genes grouped into segments irrelevant to their branches, light colors and no border: the minimal number of horizontal transfers based on DNA segments likely present in the MRCA.

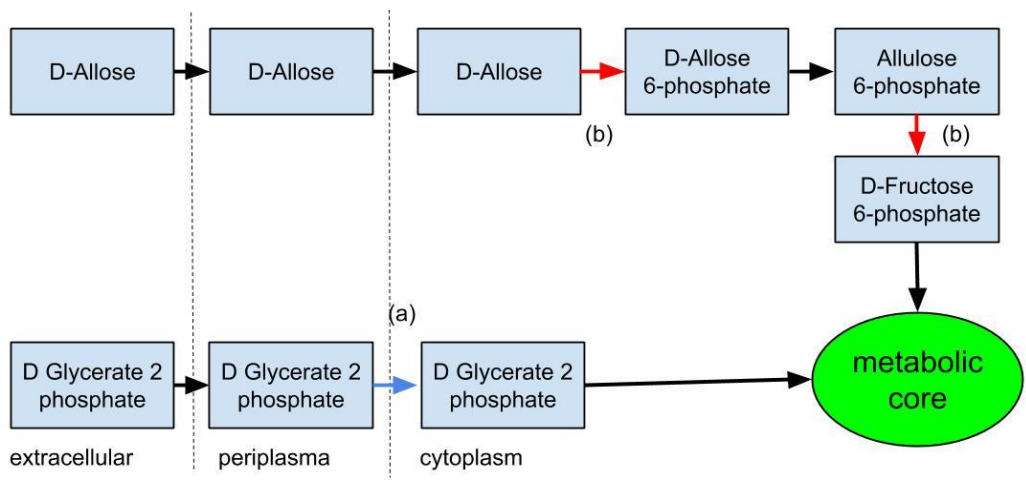

**Supplementary Figure S3**. Illustration of a complex phenotype innovation – the ability to grow in the environment corresponding to line 73,556 in **Supplementary Table S5**, distinguished by the unusual carbon source D-allose and the unusual phosphorous source D-glycerate 2 phosphate. The innovation arose in the most recent common ancestor of the APEC-O1 and S88 strains. The three relevant enzymes catalyze the reactions marked by red arrows. They were acquired in two distinct horizontal transfer events (marked by (a) and (b)) on two separate branches of the phylogeny, highlighted on the phylogenetic tree in Supplementary Figure S1.

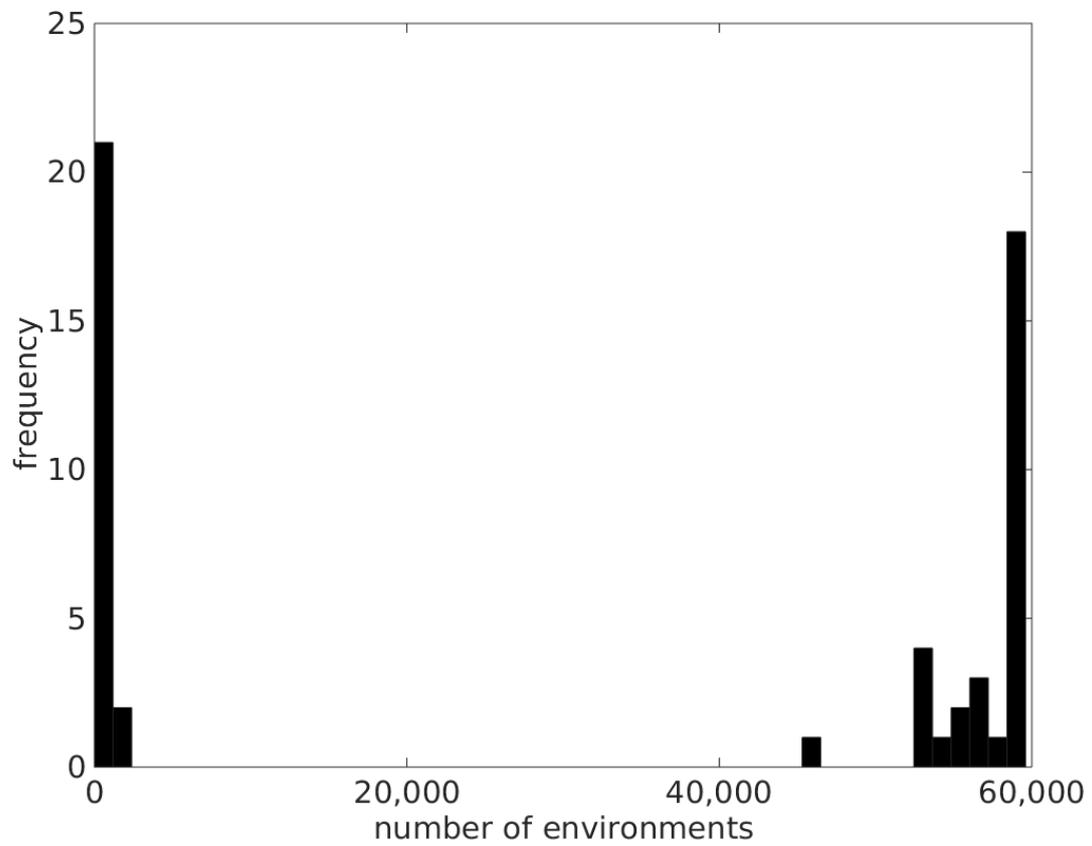

**Supplementary Figure S4**. Distribution of the number of nutritional environments in which individual extant strains can grow (including auxotrophic strains).